\documentstyle[12pt]{article}
\begin{document}
\title{STOCHASTIC CONSERVATION LAWS?}
\author{B.G. Sidharth$^*$\\ Centre for Applicable Mathematics \& Computer Sciences\\
B.M. Birla Science Centre, Hyderabad 500 063 (India)}
\date{}
\maketitle
\footnotetext{$^*$E-mail:birlasc@hd1.vsnl.net.in}
\begin{abstract}
We examine conservation laws, typically the conservation of linear momentum,
in the light of a recent successful formulation of fermions as Kerr-Newman
type Black Holes, which are created fluctuationally from a background Zero
Point Field. We conclude that these conservation laws are to be taken in the
spirit of thermodynamic laws.
\end{abstract}
Conservation Laws, as is universally known, play an important role in
Physics, starting with the simplest such laws relating to momentum and
energy. These laws provide rigid guidelines or constraints within which
physical processes take place.\\
These laws are observational, though a theoretical facade can be given by
relating them to underpinning symmetries\cite{r1}.\\
Quantum Theory, including Quantum Field Theory whilch is generally accepted as
being ultimate, is in conformity with the above picture. On the other hand the laws
of Thermodynamics have a different connotation: They are not rigid in the sense
that they are a statement about what is most likely to occur.\\
However according to a recent formulation, Quantum Theory itself takes on a
stochastic character\cite{r2,r3}. Firstly if there are $N$ particles in the universe (infact
$N \sim 10^{80}$) which has a radius $R$, then from a statistical point of
view, there is an uncertainity in the position of each particle, which as
is well known is typically taken to be a pion. This statistical uncertainity,
$l$ is given by standard theory as\cite{r4}
\begin{equation}
l \sim \frac{R}{\sqrt{N}}\label{e1}
\end{equation}
It is quite remarkable that $l$ given by equation (\ref{e1}) coincides with the
Compton wavelength of the pion.\\
Secondly according to the recent formulation which is consistent with
theory, elementary particles, infact
fermions, can be treated as Kerr-Newman type Black Holes\cite{r5}(cf.also
ref.\cite{r2} and \cite{r3}). There is ofcourse
a naked singularity but, this is blurred out by the fact that our physical
measurements are averaged over space time intervals of the order of the
Compton wavelength and time. This is the well known Zitterbewegung\cite{r6}.\\
In other words we are lead back to the idea of the chronon\cite{r7}, a minimum time unit,
supplemented by a similar minimum length interval.\\
Further, these Kerr-Newman like Black Holes are created fluctuationally from
an ambient, background, Zero Point Field (cf.ref.\cite{r2}).\\
Given the above background we consider the following simplified EPR experiment,
discussed elsewhere\cite{r8}.\\
Two structureless and spinless particles which are initially together, for
example in a bound state get separated and move in opposite directions along
the same straight line. A measurement of the momentum of one of the particles,
say $A$ gives us immediately the momentum of the other particle $B$. The
latter is equal and opposite to the former owing to the conservation law
of linear momentum. It is surprising that this statement should be true
in Quantum Theory also because the momentum of particle $B$ does not
have an apriori value, but can only be determined by a separate acausal experiment
performed on it.\\
This is the well known non locality inherent in Quantum Theory. It ceases to be mysterious
if we recognize the fact that the conservation of momentum is itself a non
local statement because it is a direct consequence of the homogeneity of
space: Infact the displacement operator $\frac{d}{dx}$ is, given the
homogeneity of space, independent of $x$ and this leads to the conservation
of momentum in Quantum Theory (cf.ref.\cite{r6}). The displacement $\delta x$ which gives rise
to the above displacement operator is an instantaneous shift of origin
corresponding to an infinite velocity and is compatible with a closed system.
It is valid if the instantaneous displacement can also be considered to be
an actual displacement in real time $\delta t$. This happens for stationary
states, when the overall energy remains constant.\\
It must be borne in mind that the space and time displacement operators are
on the same footing only in this case\cite{r9}. Indeed in relativistic Quantum
Mechanics, $x$ and $t$ are put on the same footing -  but special relativity
itself deals with inertial, that is relatively unaccelerated frames. (On the
other hand, there is not yet any successful Quantum Theory of Gravity).\\
Any field theory deals with different points at the same instant
of time. But if we are to have information about different points, then given
the finite velocity of light, we will get this information at different
times. All this information can refer to the same instant of time only in
a stationary situation. Further the field equations are obtained by a
suitable variational principle,
\begin{equation}
\delta I = 0\label{e2}
\end{equation}
In deducing these equations, the $\delta$ operator which corresponds to an
arbitrary variation, commutes with the space and time derivatives, that is
the momentum and energy operators which in our picture constitute a complete
set of observables. As such the apparently arbitrary operator $\delta$ in
(\ref{e2}) is constrained to be a function of these (stationary) variables\cite{r10}.\\
All this underscores two facts: First we implicitly consider an apriori homogenous space,
that is physical space. Secondly though we consider in the relativisitic
picture the space and time coordinates to be on the same footing, infact
they are not\cite{r11}. Our understanding or perception of the universe is based on
"all space (or as much of it as possible) at one instant of time".\\
However, in conventional theory this is at best an approximation. Moreover
in the recent formulation, as pointed out, the particles are fluctuationally created out of
a background ZPF, and, it is these $N$ particles that define physical space,
which is no longer apriori as in the Newtonian formulation. It is only in the thermodynamic limit in which
$N \to \infty$ and $l \to 0$, in (\ref{e1}), that we recover the above
classical picture of a rigid homogenous space, with the conservation laws.\\
In other words the above conservation laws are strictly valid in the
thermodynamic limit, but are otherwise approximate, though very nearly correct
because $N$ is so large. Indeed, from (\ref{e1}), we can infer that the probability
for the violation of these laws locally, is proportional to $\frac{1}{\sqrt{N}}
\sim 10^{-40}!$\\
It must be mentioned that the formulation referred to above leads to a
cosmology\cite{r12,r13}(cf.ref.\cite{r2}also) in which $\sqrt{N}$ particles
are fluctuationally created from the background ZPF (reminiscent of Prigogine's
formulation), so that the violation of energy conservation is indeed proportional
to $\frac{1}{\sqrt{N}}$. Further, this cosmological scheme is consistent not
only with known astrophysical data, but also with latest observations that the
universe is accelerating and expanding.\\
This also implies that there is a small but non-zero
probability that the measurement of the particle $A$ in the above experiment
will not give information about the particle $B$.\\
This last conclusion has also been drawn by $Gaeta$\cite{r14} who considers a background
Brownian or Nelson-Garbaczewski-Vigier noise(the ZPF referred to above) as sustaining Nelson's Stochastic
Mechanics (and the Schrodinger equation).\\
In conclusion, the conservation laws of Physics are conservation laws in the
thermodynamic sense.


\begin{thebibliography}{99}
\bibitem {r1} Roman, P., "Advanced Quantum Theory", Addison-Wesley,
Reading, Mass, 1965, p.31.
\bibitem {r2} Sidharth, B.G., Int.J. of Mod.Phys.A, 13(15), 1998. (Also
xxx.lanl.gov quant-ph 9808031).
\bibitem {r3} Sidharth, B.G., "Quantum Mechanical Black Holes:Issues and
Ramifications", Proceedings of International Symposium "Frontiers of Fundamental
Physics", Universities Press, Hyderabad, 1998. (Also xxx.lanl.gov quant-ph 9808031).
\bibitem {r4} Smolin, L., in "Quantum Concepts in space and time", Eds.,
R. Penrose, R., and  Isham, C.J., Clarendon Press, Oxford, 1986.
\bibitem {r5} Sidharth, B.G., Quantum Mechanical Black Holes:Towards a Unification
of Quantum Mechanics and General Relativity, Ind.J. Pure \& Appd.Phys., \underline{35}
(7), 1997. (Also xxx.lanl.gov quant-ph 9808020).
\bibitem {r6} Dirac, P.A.M., "The Principles of Quantum Mechanics", Clarendon
Press, Oxford, 1958.
\bibitem {r7} Caldirola, P., "Relativity, Quanta and Cosmology", Johnson
Reprint Corp., New York, 1979.
\bibitem {r8} Sidharth, B.G., "A New Approach to Locality and Causality", Vigier
Symposium, Canada, 1997. (Also xxx.lanl.gov. quant-ph 9805008).
\bibitem {r9} Davydov, A.S., "Quantum Mechanics", Pergamon Press, Oxford,
1965, p.655.
\bibitem {r10} Sidharth, B.G., Non Linear World \underline{4}, 1997.
\bibitem {r11} Misner, C.W., Thorne, K.S., and  Wheeler, J.A., "Gravitation", W.H.
Freeman, San Francisco, 1973.
\bibitem {r12} Sidharth, B.G., Intl.J.Th.Phys, 37(4), 1998.
\bibitem {r13} Sidharth, B.G., Quantum Mechanical Black Holes:An Alternative
Perspective, "Frontiers in Quantum Physics", Eds., S.C. Lim, et al, 
Springer, Singapore, 1998.
\bibitem {r14} Gaeta, G., Phys.Lett.A. 175, 267-268, 1993.
\end{thebibliography}
\end{document}